\documentclass[12pt]{article}
\usepackage{amsfonts}

\usepackage{amsmath}
\usepackage{amssymb}

\csname @addtoreset\endcsname{equation}{section}
\newcommand{\eq}{\begin{equation}}
\newcommand{\feq}{\end{equation}}
\newcommand{\eqn}{\begin{eqnarray}}
\newcommand{\feqn}{\end{eqnarray}}
\newcommand{\arr}{\begin{eqnarray*}}
\newcommand{\farr}{\end{eqnarray*}}

\font\mybb=msbm10 at 12pt
\def\bb#1{\hbox{\mybb#1}}

\def\bR {\bb{R}}

\begin{document}

\begin{titlepage}
\begin{flushright}
IFUM-684-FT\\
UTHET-01-0401\\
hep-th/0104141
\end{flushright}
\vspace{.3cm}
\begin{center}
\renewcommand{\thefootnote}{\fnsymbol{footnote}}
{\Large \bf Universality and a generalized $C$-function in CFTs \\
\vskip 3mm
with AdS Duals}
\vfill
{\large \bf {D.~Klemm$^1$\footnote{dietmar.klemm@mi.infn.it},
A.~C.~Petkou$^1$\footnote{anastasios.petkou@mi.infn.it},
G.~Siopsis$^2$\footnote{gsiopsis@utk.edu}
and D.~Zanon$^1$\footnote{daniela.zanon@mi.infn.it}}}\\
\renewcommand{\thefootnote}{\arabic{footnote}}
\setcounter{footnote}{0}
\vfill
{\normalsize
$^1$ Dipartimento di Fisica dell'Universit\`a di Milano
and\\ INFN, Sezione di Milano,
Via Celoria 16,
20133 Milano, Italy.\\
\vspace*{0.4cm}
$^2$ Department of Physics and Astronomy,\\
The University of Tennessee, Knoxville,
TN 37996 - 1200, USA.\\}
\end{center}
\vspace{2cm}

\centerline{\large \bf Abstract}
\vspace{.8cm}
\normalsize 

We argue that the thermodynamics of conformal field
theories with AdS duals exhibits a remarkable universality.
At strong coupling, a
Cardy-Verlinde entropy formula holds even when $R$-charges or bulk
supergravity scalars are turned on. In such a setting, the Casimir
entropy can be identified with a generalized $C$-function that 
changes monotonically with  
temperature as well as when non-trivial bulk scalar fields are
introduced. We
generalize the Cardy-Verlinde formula to  
cases where no subextensive part of the energy is present and further
observe that such a formula is valid for the ${\cal
N}=4$ super Yang-Mills theory in $D=4$ even at weak coupling.
Finally we show that a generalized Cardy-Verlinde  formula holds for
asymptotically flat black holes in any dimension.

\vfill

\end{titlepage}

\section{Introduction}

Recently, the thermodynamics of conformal field theories (CFTs) with
gravity duals has attracted renewed interest. Much of it
is due to the remarkable resemblance of the relevant
thermodynamical formulae to standard cosmology \cite{Verlinde}. This has led to
interesting conjectures regarding cosmological scenarios that have been
exploited in a 
number of works \cite{bulkrefs}.

Apart from cosmology the crucial observations in
\cite{Verlinde} should have implications for the thermodynamics of
CFTs {\it per se} \cite{kutasov}. 
In a previous work \cite{kps} we
made a step in that direction by studying general entropy bounds in weakly and
strongly coupled CFTs. 
The starting point of our investigation was the Cardy-Verlinde entropy formula
\begin{equation}
S = \frac{2\pi R}{n}\sqrt{E_C(2E-E_C)}\,, \label{cardyverl}
\end{equation}
for CFTs on $\bR \times S^n$, where $R$ denotes the radius of the
$S^n$ and $E_C$ is the Casimir energy defined in \cite{Verlinde}. The
above formula has been shown to hold 
in a large number of strongly coupled CFTs whose thermodynamics is
described by higher-dimensional supergravity solutions, e.~g.~ Kerr-AdS
black holes with 
one rotation parameter \cite{kps}, charged black holes in several
gauged supergravity theories \cite{cai}, or Taub-Bolt-AdS
spacetimes \cite{birm}. Whether a similar formula is 
valid for weakly coupled CFTs is not completely clear as yet.
In fact the deeper origin of (\ref{cardyverl}) remains rather obscure,
primarily because the original Cardy formula \cite{cardy}
follows from modular invariance which is 
a characteristic feature of two-dimensional conformal field theories only.

The main purpose
of the present work is to study further the above  issues. In particular,
following \cite{kps} we will argue that the
Cardy-Verlinde formula is the outcome of a striking resemblance
between the thermodynamics of CFTs with AdS duals and CFTs in
two dimensions. We will also argue that the Casimir entropy may be
viewed as a generalized $C$-function since it exhibits a
monotonic behavior under temperature 
changes as well as in cases when
a holographic  RG-flow is induced by bulk scalars in the boundary CFT. 
Such an interpretation allows us
to generalize (\ref{cardyverl}) to the case where no
subextensive part of the energy is present, i.~e.~when the Casimir
energy vanishes but the Casimir entropy is different from
zero.
Moreover we show that (\ref{cardyverl}) is valid also
for weakly coupled $D=4$,
${\cal N}=4$ super Yang-Mills theory, with an overall coefficient
$4\pi R/3$ instead of $2\pi R/3$.

The paper is organized as follows. In Section 2 we study the
thermodynamics of black hole solutions in gauged supergravities with
non-trivial bulk scalar fields. In all cases considered we find that
(\ref{cardyverl}) is valid for the corresponding dual CFTs. In Section
3 we discuss the validity of (\ref{cardyverl}) for CFTs in flat
spaces, e.~g.~dual to black holes with flat horizons. This allows us
to discuss the monotonicity properties of the Casimir entropy under
temperature changes as well as under the switching on of bulk scalars.
Our results suggest that the Casimir entropy can be interpreted as a
monotonic, generalized $C$-function for theories at finite temperature.
In Section 4 we show that the Cardy-Verlinde formula (\ref{cardyverl})
remarkably holds also for the weakly coupled ${\cal N}=4$ SYM in
$D=4$, with an overall coefficient which is twice the one in the
strongly coupled limit.
Finally, in Section 5 we show that a generalized Cardy formula
holds also for asymptotically flat black holes in any dimension.
We conclude and present an outlook of our ideas in Section 6.

\section{Black holes in gauged supergravity and the Cardy-Verlinde
entropy formula}

We begin by considering charged black holes in diverse gauged
supergravities studied in \cite{cai}. The $STU$ model of $D=5$, $N=2$
gauged supergravity\footnote{This model can also be embedded into $D=5$, $N=8$
gauged supergravity.} admits the black hole solutions \cite{wafic,mirjam}
\begin{equation}
ds^2 =-(H_1H_2H_3)^{-2/3} fdt^2 +(H_1H_2H_3)^{1/3}(f^{-1}dr^2 +r^2d\Omega_3^2),
      \label{bhstu}
\end{equation}
where 
\begin{equation}
f = 1 - \frac{\mu}{r^2} + r^2R^{-2} H_1 H_2 H_3, \quad  H_I=1+\frac{q_I}{r^2},
\quad I=1,2,3. \label{harm}
\end{equation}
The moduli $X^I$ and the gauge potentials $A^I$ are given by
\begin{equation}
X^I = H_I^{-1}(H_1H_2 H_3)^{1/3}, \quad A^I_t = \frac{\tilde{q}_I}{r^2 +q_I},
\end{equation}
where the $\tilde{q}_I$ denote the physical charges related to the $q_I$ by
\begin{equation}
q_I=\mu \sinh^2\beta_I, \quad \tilde q_I =\mu \sinh\beta_I \cosh\beta_I.
\end{equation}
The BPS limit \cite{ali} is reached when the nonextremality parameter
$\mu$ goes to zero and $\beta_I \to \infty$, with $q_I$ fixed\footnote{The
reader should note that the BPS limit represents naked singularities.}.
The horizon coordinate $r_+$ obeys
\begin{equation}
\mu = r_+^2 \left (1+\frac{1}{R^2 r_+^4} \prod _{I=1}^3 (r_+^2 + q_I) \right).
\end{equation}
The black hole mass and the Bekenstein-Hawking entropy read, respectively
\begin{eqnarray}
M &=& \frac{\pi}{4G}\left (\frac{3}{2}\mu +\sum_{I}q_I\right), \nonumber \\
S &=& \frac{\pi^2}{2G}\sqrt{\prod_I (r_+^2 + q_I)}. \label{S5d}
\end{eqnarray}
We define the excitation energy above the BPS state,
\begin{equation}
E = M - M_{BPS} = \frac{3\pi}{8G}\mu = \frac{3\pi}{8G}r_+^2\left[1 +
                  \frac{1}{R^2r_+^4}\prod_I(r_+^2 + q_I)\right].
\label{E5d}
\end{equation}
Following \cite{kps} we define a parameter $\Delta$ by
\begin{equation}
\Delta^{-2} = \frac{1}{R^2r_+^4}\prod_I(r_+^2 + q_I). \label{Delta5D}
\end{equation}
This yields
\begin{equation}
2ER = \frac{3}{2\pi} S\Delta [1 + \Delta^{-2}], \label{ER5d}
\end{equation}
which is exactly the behavior of a two-dimensional CFT with characteristic
scale $R$, temperature $\tilde{T} = 1/(2\pi R\Delta)$, and central charge
proportional to $S\Delta$. This resemblance motivates to define the
Casimir energy as the subextensive part of (\ref{ER5d}) 
\begin{equation}
E_CR = \frac{3}{2\pi}S\Delta. \label{EC5d}
\end{equation}
Now one easily verifies that the quantities (\ref{S5d}), (\ref{E5d}) and
(\ref{EC5d}) satisfy exactly the Cardy-Verlinde formula (\ref{cardyverl})
for $n=3$. As in \cite{Verlinde} we can define the Casimir entropy
$S_C$ by
\begin{equation}
S_C = \frac{2\pi}{n}E_CR = S\Delta.
\end{equation}
In terms of $S_C$, (\ref{ER5d}) can be rewritten as
\begin{equation}
2ER = \frac{3}{2\pi} S_C [1 + \Delta^{-2}],
\end{equation}
which allows to interpret the Casimir entropy
as a generalization of 
the central charge to higher dimensions. This interpretation of $S_C$
was already pointed out by Verlinde \cite{Verlinde}.\\

The above considerations can be easily generalized to black holes
in $D=4$ and $D=7$ gauged supergravities as well. That (\ref{cardyverl})
holds in these cases was shown in \cite{cai}, but we would like
to reformulate the results of \cite{cai} in a different way, in order to
stress that the validity of (\ref{cardyverl}) can be traced
back to an effective two-dimensional behavior.\\
Black hole solutions in a truncated version of $D=4$, $N=8$ gauged
supergravity have been found in \cite{duff}. The metric is given by
\begin{equation}
ds^2 = -(H_1H_2H_3H_4)^{-1/2} fdt^2 +(H_1H_2H_3H_4)^{1/2}(f^{-1}dr^2
   +r^2 d\Omega_2^2),
\end{equation}
where
\begin{equation}
f = 1-\frac{\mu}{r} +R^{-2}r^2 \prod_{I=1}^4H_I, \quad H_I =1+\frac{q_I}{r},
\quad I=1,\ldots,4.
\end{equation}
Furthermore there are gauge fields and scalars turned on
(cf.~\cite{duff} for details). The nonextremality parameter $\mu$ is
given in terms of the horizon radius $r_+$ by
\begin{equation}
\mu = r_+\left( 1+\frac{1}{l^2 r_+^2} \prod_I (r_+ + q_I) \right).
\end{equation}
The mass and entropy read
\begin{eqnarray}
M &=& \frac{1}{4G}(2\mu + \sum_I q_I), \nonumber \\
S &=& \frac{\pi}{G}\sqrt{\prod_I (r_+ + q_I)}, 
\end{eqnarray}
respectively. Again we define the excitation energy $E=M-M_{BPS}$
above the BPS state (which is the one with $\mu = 0$), yielding
\begin{equation}
2ER = \frac{r_+R}{G}\left[1 + \frac{1}{R^2r_+^2}\prod_I (r_+ + q_I)\right].
\end{equation}
Defining also
\begin{equation}
\Delta^{-2} = \frac{1}{R^2r_+^2}\prod_I (r_+ + q_I),
\end{equation}
we obtain
\begin{equation}
2ER = \frac{1}{\pi} S\Delta [1 + \Delta^{-2}], \label{ER4d}
\end{equation}
which resembles again the behavior of a two-dimensional CFT.
The Casimir energy is then easily determined from (\ref{ER4d}) as
the subextensive part, i.~e.
\begin{equation}
E_CR = \frac{1}{\pi}S\Delta = \frac{1}{\pi}S_C.
\end{equation}
The quantities $S$, $E$ and $E_C$ satisfy again the Cardy-Verlinde
formula (\ref{cardyverl}) for $n=2$.\\

Two-charge black hole solutions of $D=7$, $N=4$ gauged supergravity
can be found in \cite{liu,mirjam}. We give here only the metric,
which reads
\begin{equation}
ds^2 = -(H_1H_2)^{-4/5} fdt^2 +(H_1H_2)^{1/5}(f^{-1}dr^2 +r^2d\Omega_5^2),
\end{equation}
with
\begin{equation}
f(r) = 1-\frac{\mu}{r^4} +r^2R^{-2}H_1 H_2, \quad H_I =1+\frac{q_I}{r^4},
\quad I = 1,2.
\end{equation}
The nonextremality parameter $\mu$ is related to the black hole horizon
$r_+$ by
\begin{equation}
\mu = r_+^4 +\frac{1}{r_+^2 l^2}\prod_I (r_+^4 + q_I).
\end{equation}
The mass and entropy are
\begin{eqnarray}
M &=& \frac{\pi^2}{4G}\left (\frac{5}{4}\mu +\sum_I q_I\right), \nonumber \\
S &=& \frac{\pi^3 r_+}{4G} \sqrt{\prod_I (r_+^4 + q_I)}. \nonumber \\
\end{eqnarray}
Defining as above
\begin{equation}
E = M - M_{BPS} = M - M_{\mu = 0}\,,
\end{equation}
and
\begin{equation}
\Delta^{-2} = \frac{1}{R^2r_+^6}\prod_I(r_+^4 + q_I),
\end{equation}
one obtains
\begin{equation}
2ER = \frac{5}{2\pi}S\Delta [1 + \Delta^{-2}] =
      \frac{5}{2\pi}S_C[1 + \Delta^{-2}],
\end{equation}
so that $E_CR = (5/2\pi)S_C$, and (\ref{cardyverl}) still holds.\\

The above considerations go through also
for Reissner-Nordstr\"om-AdS black holes in arbitrary dimension $n+2$,
\begin{eqnarray}
ds^2 &=& -H^{-2}fdt^2 + H^{\frac{2}{n-1}}(f^{-1}dr^2 + r^2 d\Omega_n^2),
         \nonumber \\
H &=& 1 + \frac{q}{r^{n-1}}, \nonumber \\
f &=& 1 - \frac{\mu}{r^{n-1}} + r^2R^{-2}H^{\frac{2n}{n-1}}, \nonumber
\end{eqnarray}
with mass and entropy given by
\begin{eqnarray}
M &=& \frac{nV_n}{16\pi G}(\mu + 2q), \nonumber \\
S &=& \frac{V_n}{4G}(r_+^{n-1} + q)^{\frac{n}{n-1}}, \nonumber
\end{eqnarray}
where $V_n$ denotes the volume of the unit $n$-sphere.
With $E = M - M_{\mu=0}$ one finds
\begin{equation}
2ER = \frac{n}{2\pi}S_C[1 + \Delta^{-2}], \label{ERgen}
\end{equation}
where
\begin{equation}
\Delta^{-2} = \frac{1}{R^2r_+^{2(n-1)}}(r_+^{n-1} + q)^{\frac{2n}{n-1}},
\end{equation}
and the central charge $S_C = S\Delta$. This leads
again to (\ref{cardyverl}).\\

In order to emphasize the complete universality of (\ref{cardyverl}) and
(\ref{ERgen}), let us consider as a final example the Kerr-AdS black hole
in five dimensions with two rotation parameters. The metric
reads \cite{hawking}
\begin{eqnarray}
ds^2 &=& - \frac{\Delta_r}{\rho^2} (dt - \frac{a \sin^2\theta}{\Xi_a}d\phi -
\frac{b \cos^2\theta}{\Xi_b} d\psi)^2 +
\frac{\Delta_{\theta}\sin^2\theta}{\rho^2} (a dt -
\frac{(r^2+a^2)}{\Xi_a} d\phi)^2 \nonumber \\
&& + \frac{\Delta_{\theta}\cos^2\theta}{\rho^2} (b dt -
\frac{(r^2+b^2)}{\Xi_b} d\psi)^2 + \frac{\rho^2}{\Delta_r} dr^2 +
\frac{\rho^2}{\Delta_{\theta}} d\theta^2 \label{metr2rot} \\
&& + \frac{(1+r^2 R^{-2})}{r^2 \rho^2}
\left ( ab dt - \frac{b (r^2+a^2) \sin^2\theta}{\Xi_a}d\phi - \frac{a
    (r^2 + b^2) \cos^2 \theta}{\Xi_b} d\psi \right )^2, \nonumber
\end{eqnarray}
where 
\begin{eqnarray}
\Delta_r &=& \frac{1}{r^2} (r^2 + a^2) (r^2 + b^2) (1 + r^2 R^{-2}) - 2m;
\nonumber \\
\Delta_{\theta} &=& \left ( 1 - a^2 R^{-2} \cos^2\theta - b^2 R^{-2}
  \sin^2\theta \right ); \\
\rho^2 &=& \left ( r^2 + a^2 \cos^2\theta + b^2 \sin^2\theta \right);
\nonumber \\
\Xi_a &=& (1-a^2 R^{-2}); \hspace{5mm} \Xi_b = (1 -b^2 R^{-2}). \nonumber 
\end{eqnarray}
The horizon location $r_+$ is the largest root of $\Delta_r = 0$.
The mass and the entropy are given by \cite{hawking}
\begin{eqnarray}
M &=& \frac{3\pi}{8\Xi_a\Xi_bGr_+^2}(r_+^2 + a^2)(r_+^2 + b^2)
      (1 + r_+^2R^{-2}), \nonumber \\
S &=& \frac{\pi^2}{2 r_{+} \Xi_a \Xi_{b}G} (r_{+}^2 + a^2)(r_{+}^2 + b^2).
      \nonumber
\end{eqnarray}
With $E=M$, $\Delta^{-1} = r_+/R$ and $S_C = S\Delta$ we can write
\begin{equation}
2ER = \frac{3}{2\pi}S_C [1 + \Delta^{-2}],
\end{equation}
and thus also in the rotating case the two-dimensional CFT behavior
persists, and the Cardy-Verlinde formula (\ref{cardyverl}) is satisfied.\\

\section{Universality and monotonicity properties of the generalized
$C$-function} 

The results obtained in the previous Section reveal a remarkable
universality in the thermodynamics of CFTs with AdS duals. The
supergravity theories studied above contain bulk gauge fields that couple
to $R$-currents on the 
boundary and therefore the dual CFTs have $R$-charges turned 
on. Also, they contain bulk scalar fields that either couple to
operators in the 
boundary CFT which acquire nonvanishing expectation values, or else 
induce RG-flows \cite{GirFreed}.
In all cases  we found  a striking resemblance
of the thermodynamics of the boundary CFT to the one of a CFT in two
dimensions. The only 
parameter that apparently encodes the properties of the CFT under
consideration is 
$\Delta$. The same universal 
behavior is observed also for rotating black holes
which are dual to strongly coupled CFTs on a rotating Einstein
universe \cite{kps}. 

An important ingredient in our calculations has been the fact
that we were able to define the Casimir entropy $S_C$ in such a way
that it closely resembles a two-dimensional central charge. We would
like 
to substantiate this observation.  
At high temperatures one easily shows that $S_C$ is proportional  
to the derivative of the entropy with respect
to the temperature. This means that it is related to the number of
degrees of freedom between $T$ and $T + dT$, hence it lends itself as
a good candidate 
for a generalized $C$-function in theories at finite temperature. For
such an interpretation to be meaningful, one should at least demonstrate that
$S_C$  possesses certain monotonicity 
properties. For the general cases considered in Section 2,
$S_C$ is a function of {\it both} the temperature and the
$R$-charges. Then it is 
crucial to study the behavior of $S_C$ 
{\it both} under temperature changes as well as when non-trivial bulk fields
are turned on. The latter case should be related to the behavior of the
Casimir entropy under RG-flows, in much the same way as the presence
of non-trivial fields in the bulk theory 
induces a holographic RG-flow in the boundary theory. 

In order to analyze the behavior of the Casimir entropy under
temperature changes we  
consider the field theory dual to the Schwarz\-schild-AdS
black hole in $n+2$ dimensions. In this case one has \cite{kps}
\begin{equation}
S_C = \frac{V_nR}{4G}r_+^{n-1}.
\end{equation}
One easily finds that the monotonicity property
\begin{equation}
T\frac{dS_C}{dT} \ge 0,
\end{equation}
is equivalent to $r_+^2/R^2 \ge (n-1)/(n+1)$, and thus coincides
with the region of stability above the Hawking-Page phase
transition \cite{hp} where the free energy is a concave function
of the temperature. This indicates that there is a direct
relationship between monotonicity of the generalized $C$-function
and local thermodynamic stability.

Next we want to study how $S_C$ changes when 
non-trivial bulk 
scalars are turned on. In many cases, turning on bulk scalar has been
shown to induce an RG-flow in the dual theory \cite{GirFreed}. Since such
holographic RG-flows have been 
studied for dual field theories that
live in flat space, the first step is to understand how to
generalize the Cardy-Verlinde formula and the definition of the Casimir
entropy for such cases, e.g. when no
subextensive part  of the energy is present. 
Let us consider the AdS black holes in dimension $n+2$ with flat horizon
and metric given by
\begin{eqnarray}
ds^2 &=& -f(r)dt^2 + f^{-1}(r)dr^2 + r^2d\vec{x}^2, \label{flatBH} \\
f(r) &=& -\frac{2m}{r^{n-1}} + \frac{r^2}{R^2}, \nonumber
\end{eqnarray}
where $d\vec{x}^2$ denotes the line element on a flat $n$-dimensional
Euclidean space. The mass and entropy densities read
\begin{eqnarray}
\frac EV &=& \frac{n}{16\pi G}\frac{r_+^{n+1}}{R^2}, \nonumber \\
\frac SV &=& \frac{r_+^n}{4G}, \nonumber
\end{eqnarray}
where $r_+$ is the largest root of $f(r) = 0$.
In two-dimensional conformal field theories the Cardy formula
\begin{equation}
S = 2\pi \sqrt{\frac{cL_0}{6}} \label{cardy}
\end{equation}
is valid only when the conformal weight of the ground state is zero,
otherwise (in the presence of a Casimir energy) one has to modify
(\ref{cardy}) to
\begin{equation}
S = 2\pi \sqrt{\frac c6\left(L_0 - \frac{c}{24}\right)}, \label{cardymod}
\end{equation}
which is the two-dimensional form of (\ref{cardyverl}).
For the black holes (\ref{flatBH}), the dual CFT lives on a flat space,
and thus the energy has no subextensive part. Since the Casimir energy
vanishes, the Cardy-Verlinde formula
(\ref{cardyverl}) makes no sense in this case. However, we saw above
that the appropriate generalization of the central charge $C$ to
higher dimensions is essentially given by the Casimir entropy.
If we use $E_CR = (n/2\pi)S_C$ in (\ref{cardyverl}), and drop the
subtraction of $E_C$ in analogy with (\ref{cardy}), we obtain
the generalization of (\ref{cardy}) to $n+1$ dimensions,
\begin{equation}
S = \frac{2\pi}{n}\sqrt{\frac{cL_0}{6}}, \label{cardyverlmod}
\end{equation}
where $L_0 = ER$ and $c/6 = nS_C/\pi$.
Using $\Delta^{-1} = r_+/R$, one easily verifies that the
black holes (\ref{flatBH}) satisfy the modified Cardy-Verlinde formula
(\ref{cardyverlmod}). 

Now we are ready to proceed. A suitable class of models that exhibit both holographic RG-flows and
non-zero temperature are the $STU$ black holes (\ref{bhstu}) with flat
horizons \cite{wafic,mirjam}, 
given by\footnote{Note that this solution can be obtained from (\ref{bhstu})
by writing the metric on $S^3$ as $d\Omega_3^2 = d\chi^2 + \sin^2\chi
d\Omega_2^2$, and then performing the scaling limit $\chi \to \epsilon\chi$,
$r \to r/\epsilon$, $t \to \epsilon t$, $\mu \to \mu/\epsilon^4$,
$\beta_I \to \epsilon\beta_I$, $\epsilon \to 0$.}
\begin{eqnarray}
\label{STUflat}
ds^2 &=& -(H_1H_2H_3)^{-2/3}f dt^2 + (H_1H_2H_3)^{1/3}(f^{-1} dr^2 +
         r^2 d\vec{x}^2), \\
X^I &=& H_I^{-1}(H_1H_2H_3)^{1/3}, \qquad A^I_t =
        \frac{\sqrt{q_I\mu}}{r^2 + q_I}, \nonumber
\end{eqnarray}
where 
\begin{equation}
f = - \frac{\mu}{r^2} + r^2R^{-2} H_1 H_2 H_3\,,
\end{equation}
and the harmonic functions are given in (\ref{harm}). 
In \cite{Behrndt} it has been shown  that, in the extremal limit, the presence
of non-trivial\footnote{It was shown in \cite{Behrndt} that one
requires $q_1\neq q_2\neq q_3$ for non-singular flows.} bulk 
scalar fields gives rise to a holographic RG-flow. Indeed, when $\mu=0$
the metric (\ref{STUflat}) can be put in the domain wall form
\begin{equation}
\label{DW}
ds^2 = e^{2A(\rho)}[-dt^2 +d\vec{x}^2] +d\rho^2\,,
\end{equation}
with
\begin{equation}
e^{2A(\rho)} = \frac{r^2}{R^2}(H_1H_2H_3)^{\frac 13}
\,,\qquad \frac{dr}{d\rho} =
\frac{r}{R}(H_1H_2H_3)^{\frac{1}{3}}\,.
\end{equation}
Then one computes the $C$-function \cite{GirFreed} which describes the
induced RG-flow 
as\footnote{Here the 5-dimensional Newton constant $G$ satisfies the
AdS/CFT relation $\frac{2N^2}{\pi}= \frac{R^3}{G}$.}
\begin{equation}
\label{CGir}
C \sim \frac{1}{G[A'(\rho)]^3} =
\frac{(3R)^3}{G}\frac{(H_1H_2H_3)^2}{(H_1H_2 + H_1H_3+H_2H_3)^3}\,.
\end{equation}
The UV limit is reached when $r\rightarrow \infty$ and the IR limit
when $r\rightarrow 0$. Equivalently, the UV limit is reached as
$q_1,q_2,q_3\rightarrow 0$ and the IR limit as $q_1,q_2,q_3\rightarrow
\infty$. The corresponding expressions for the $C$-function read then
\begin{equation}
\label{CUVIR}
C^{UV} \sim\frac{R^3}{G} > C^{IR} \sim
\frac{(3R)^3}{G}\frac{(q_1q_2q_3)^2}{ (q_1q_2+q_1q_3+q_2q_3)^3}\,.
\end{equation}
In the example above, the extremal limit leads to a field theory at zero temperature and one obtains the usual generalized $C$-function defined from the holographic RG-flow of theories which admit dual gravitational descriptions.

Away from the extremal limit we obtain a dual field theory at finite temperature. We want now to show that also at finite temperature we can implement an RG-flow. We argue as follows.
For the metric (\ref{STUflat}), the 
mass and Bekenstein-Hawking entropy are
\begin{eqnarray}
E &=& \frac{3V_3}{16\pi G}\mu = \frac{3V_3}{16\pi G}\frac{1}{R^2r_+^2}
      \prod_I(r_+^2 + q_I)\,, \label{ESTUflat} \\
S &=& \frac{V_3}{4G}\sqrt{\prod_I(r_+^2 + q_I)}\,. \label{SSTUflat}
\end{eqnarray}
These are interpreted as the mass and the entropy of the corresponding
dual CFT. With $\Delta^{-2}$ defined in (\ref{Delta5D}), one gets for
the Casimir entropy
\begin{equation}
S_C = S\Delta = \frac{V_3Rr_+^2}{4G}. \label{SCSTUflat}
\end{equation}
One easily checks that (\ref{ESTUflat}), (\ref{SSTUflat}) and
(\ref{SCSTUflat}) satisfy (\ref{cardyverlmod}) with $L_0 = ER$
and $c/6 = 3S_C/\pi$.
Notice now that since the horizon coordinate obeys
\begin{equation}
\label{horSTUflat}
\mu = \frac{1}{R^2 r_+^2} \prod _{I=1}^3 (r_+^2 + q_I) \,,
\end{equation}
the value of $S_C$ in (\ref{SCSTUflat}) depends implicitly on the
charges $q_I$. Following up with our proposal to interpret $S_C$ as a
generalized $C$-function, we keep the effective
temperature of the dual CFT fixed and allow only the variation of the
$q_I$. Now, as we have done in the zero temperature limit, we want to
see how $S_C$ changes in the two limits $q_I\rightarrow 0,\infty$,
which were the UV and IR limits respectively of the CFT dual to the 
extremal case.  Then, by virtue of (\ref{horSTUflat}) we see that
under such a variation $S_C$ changes monotonically. 
More specifically in the UV limit where $q_I\rightarrow 0$, which is equivalent to taking $r_+$ large, we obtain the usual result for the Casimir
entropy  of ${\cal N}=4$ SYM at strong coupling
\begin{equation}
S_C^{UV} =\frac{V_3R^3}{4G}\left(\frac{r^0_+}{R}\right)^2\,,
\end{equation}
where now $r_+^0$ denotes the solution of (\ref{horSTUflat}) for
$q_I=0$. 
In the IR limit where $q_I\rightarrow \infty$, or equivalently
for small $r_+$ ($r^2_+\ll q_I$), we obtain
\begin{equation}
\label{SCIR}
S_C^{IR} = S_C^{UV} \frac{r_+^3}{\sqrt{q_1q_2q_3}}\ll S_C^{UV}\,.
\end{equation}

%

Using the same approach as above, we moreover find a quite interesting
RG-flow for a boundary CFT that exhibits a temperature phase 
transition. This is the case of the CFT dual to the STU black hole
with spherical horizons. It is described by the metric 
\begin{equation}
\label{STUspherical}
ds^2 = -(H_1H_2H_3)^{-2/3}f dt^2 + (H_1H_2H_3)^{1/3}(f^{-1} dr^2 +
         r^2 d\Omega_3^2)\,,
\end{equation}
where 
\begin{equation}
\label{horSTUsphere}
f = 1- \frac{\mu}{r^2} + r^2R^{-2} H_1 H_2 H_3\,.
\end{equation}
The horizon
coordinate now obeys 
\begin{equation}
\label{horSTUfl}
\mu = r_+^2\left( 1+\frac{1}{R^2 r_+^4} \prod _{I=1}^3 (r_+^2 + q_I)\right) \,,
\end{equation}
Notice that a solution exists only if
\begin{equation}
\mu R^2 > q_1q_2 + q_2q_3 + q_3q_1\,.
\end{equation}
To relate our discussion with the RG-flow described before, we wish to
consider the extremal limit $\mu\to 0$. This, however, can only be
achieved if at 
the same time we let $R\to \infty$. 
From (\ref{horSTUsphere}) it is natural to take 
\begin{equation}
\mu \gtrsim \frac{\sqrt{q_1q_2q_3}}{R}\,.
\end{equation}
The Casimir entropy is still given by (\ref{SCSTUflat}).
An important property of the theory dual to (\ref{STUspherical}) is
that it exhibits a Hawking-Page phase transition at a temperature
defined by the relation  $\Delta = 1$. This is exactly the point where
the following entropy bound
\begin{equation}
\frac{S}{S_B} = \frac{2\Delta^{-1}}{1+\Delta^{-2}} \le
1\,,\,\,\,\,\,\,\,\, S_B = \frac{2\pi ER}{3}\,,
\end{equation}
is saturated. As discussed before, at this special point the flow of
$S_C$ induced by the non-trivial 
moduli is monotonic. Specifically, for large $r_+$ (large black hole), 
we have $r_+ = R$ due to $R^2\gg q_I$. Therefore we obtain
\begin{equation}
S_C = S = \frac{\pi^2}{2G} \; R^3\,,
\end{equation}
which corresponds to the ${\cal N}=4$ Super Yang-Mills model.
For small $r_+$, in which case
$Rr_+^2 = \sqrt{q_1q_2q_3}$, we find
\begin{equation}
\label{SCBPS}
S_C = \frac{\pi^2}{2G} \; \sqrt{q_1q_2q_3}\,.
\end{equation}
We can now see that (\ref{SCBPS}) corresponds to a BPS state. 
This can be obtained from a D1-brane of charge $Q_1
\sim q_1/\sqrt G$, together 
with a D5-brane of charge $Q_5 \sim q_2/\sqrt G$ in six-dimensions
\cite{Youm}. By
compactifying the sixth 
dimension along a circle of radius $R_0$, we obtain momenta $P\sim
N/R_0$, where 
$N\sim q_3/G$. Then a Kaluza-Klein reduction produces a black hole
with charges $Q_1$ $Q_5$ and $N$, and entropy 
\begin{equation}
S \sim \sqrt{Q_1Q_5N} \sim {\frac{1}{G}}  \,\sqrt{q_1q_2q_3}
\end{equation}

The examples studied above suggest that indeed the Casimir entropy $S_C$
is a natural candidate for a generalized $C$-function for field theories
that have a dual supergravity description, both at zero and at finite
temperature. In general $S_C$ is a function of {\it both} the
temperature and 
the 
renormalization scale (which enters the game through the warp-factor
$A(\rho)$ in (\ref{DW})) of the field theory. We have found that $S_C$
is a monotonic function in each variable, while keeping the other one
fixed. 
At fixed energy scale we move in the space of conformally invariant theories
at varying temperature, while at fixed temperature we move along the renormalization group trajectories. 

\section{Free field theory side}

Now we go back to manifolds
of the form $\bR \times S^n$, and start with
a system of $N_B$ scalars, $N_F$ Weyl fermions and $N_V$ vectors
on $\bR \times S^3$, with the radius of the $S^3$ given by $R$.
The free energy has been computed in \cite{constable,kutasov}, and reads
\begin{equation}
-FR = a_4\delta^{-4} + a_2\delta^{-2} + a_0,
\end{equation}
where $\delta^{-1} = 2\pi RT$ and
\begin{eqnarray}
a_4 &=& \frac{1}{720}(N_B + \frac 74 N_F + 2N_V), \quad
a_2 = -\frac{1}{24}(\frac 14 N_F + 2N_V), \nonumber \\
a_0 &=& \frac{1}{240}(N_B + \frac{17}{4}N_F + 22N_V),
\end{eqnarray}
satisfying the constraint $3a_4 = a_2 + a_0$. The entropy and energy are,
respectively,
\begin{equation}
S = 2\pi(4a_4\delta^{-3} + 2a_2\delta^{-1}), \quad
ER = 3a_4\delta^{-4} + a_2\delta^{-2} - a_0.
\end{equation}
One thus obtains
\begin{equation}
2ER = \frac{1}{2\pi}S\delta [1 + \delta^{-2}]
      \frac{3a_4 - a_0\delta^2}{2a_4 + a_2\delta^2}.
\end{equation}
Remarkably, for the ${\cal N}=4$ SYM model, this simplifies to
\begin{equation}
2ER = \frac{3}{4\pi}S\delta [1 + \delta^{-2}], \label{ERSYM}
\end{equation}
which resembles again the behavior of a two-dimensional CFT with
central charge given by $S_C \equiv S\delta$. We can define the
Casimir energy as the subextensive part of (\ref{ERSYM}), i.~e.
\begin{equation}
E_CR = \frac{3}{4\pi}S\delta. \label{ECfreefields}
\end{equation}
It is then easily shown that a Cardy-Verlinde formula
\begin{equation}
S = \frac{4\pi R}{3}\sqrt{E_C(2E - E_C)} \label{cardyverlweak}
\end{equation}
holds. Note that the prefactor in (\ref{cardyverlweak}) is twice the
one in (\ref{cardyverl}), which is valid at strong `t Hooft coupling.
Note also that the parameter $\delta$ at weak coupling is different
from $\Delta$ used at strong coupling. Whereas $\delta^{-1}$ is related to
the true temperature, $\Delta^{-1}$ represents an effective
temperature \cite{kps}, which, for $T \to \infty$, becomes again
directly related to $T$.

We thus found that for the ${\cal N}=4$ SYM model on $\bR\times S^3$,
a generalized Cardy formula

\begin{equation}
S = \frac{b \pi R}{3}\sqrt{E_C(2E-E_C)} \label{cvgeneral}
\end{equation}
holds, with $b=2$ for strong coupling and $b=4$ for free fields.
The fact that (\ref{cvgeneral})
is valid in both extremal regimes suggests that it may hold for
every value of the coupling, with $b$ being a function of $g_{YM}^2N$,
that interpolates smoothly (at least for temperatures above the Hawking-Page
phase transition)
between the values $2$ for $g_{YM}^2N\to\infty$
and $4$ for $g_{YM}^2N\to 0$. As a first step to see whether such
a conjecture makes sense, one would wish to compute the leading
stringy corrections
to the supergravity approximation, in order to see if they
are indeed positive. In the low energy effective action,
massive string modes manifest themselves
as higher derivative curvature terms. In type IIB supergravity, the
lowest correction is of order ${\alpha'}^3R^4$ \cite{daniela}, where
$R$ denotes the Riemann tensor. The leading stringy corrections
to the thermodynamics of Schwarz\-schild-AdS black holes (without
rotation or charges) have been computed
in \cite{Landsteiner:1999gb,Caldarelli:1999ar},
generalizing the calculations for SYM on flat
space \cite{Gubser:1998nz,Pawelczyk:1998pb}.
As we said, we expect the relation
\begin{equation}
2ER = \frac{3}{b\pi}S\Delta [1 + \Delta^{-2}] \label{universalrel}
\end{equation}
to be completely universal, i.~e.~to hold also for the
${\alpha'}^3$-corrected thermodynamical quantities. The problem is
now that in (\ref{universalrel}) one has two unknown functions,
namely $b$ and $\Delta$. In fact, for the free field theory
$\Delta^{-1} = \delta^{-1} = 2\pi RT$, whereas for strong coupling
$\Delta^{-1} = r_+/R$ is a complicated function of the temperature.
It is thus clear that also $\Delta$ changes with
the coupling. One might be tempted to set $\tilde{\Delta} = R/\tilde{r}_+$
for the ${\alpha'}^3$-corrected function, where $\tilde{r}_+$ is the 
corrected horizon location, which is known \cite{Caldarelli:1999ar}. This
choice is however not obvious, because in principle  $\tilde{\Delta}$
could be a more complicated expression in terms of $\tilde{r}_+$.
Note also that at strong coupling one has $E_C = (n+1)E - nTS$ with
$n=3$, whereas for free fields, (\ref{ECfreefields}) satisfies
$E_C = (4E - 3TS)/2$. We will leave the further study of stringy
corrections to the Cardy-Verlinde formula for a future publication.

\section{A generalized Cardy-Verlinde formula for asymptotically flat
black holes}

All cases where the Cardy-Verlinde formula has been
shown to hold up to now had as a necessary ingredient
a negative cosmological constant, or, more generally,
a certain potential term for supergravity scalars. This guarantees
that the theory admits AdS vacua, and thus one has a
dual description in terms of a conformal field theory
on the boundary of AdS. A natural question is whether
the Cardy-Verlinde formula holds in a more general
setting, e.~g.~for black holes that are asymptotically
flat rather than approaching AdS space. We will show
that this is indeed the case. Let us first consider
the Schwarz\-schild solution in $n+2$ dimensions,
given by
\begin{equation}
ds^2 = -\left(1 - \frac{2m}{r^{n-1}}\right)dt^2 +
       \left(1 - \frac{2m}{r^{n-1}}\right)^{-1}dr^2 + r^2 d\Omega_n^2.
\label{schwarz}
\end{equation}
The mass, entropy and temperature read
\begin{equation}
E = \frac{nV_n}{16\pi G}r_+^{n-1}, \qquad
S = \frac{V_n}{4G}r_+^n, \qquad
T = \frac{n-1}{4\pi r_+}, \label{S}
\end{equation}
where $r_+$ is the horizon location obeying $r_+^{n-1} = 2m$.
As in \cite{Verlinde}, we can now define the Casimir energy 
as the violation of the Euler identity,
\begin{equation}
E_C = n(E - TS + pV). \label{EC}
\end{equation}
Let us assume for the moment that, like their Schwarz\-schild-AdS
counterparts, also the black holes (\ref{schwarz}) in flat space
are described by a conformal field theory in $n+1$ dimensions.
Then the stress tensor is traceless,
which implies the equation of state $pV = E/n$. Using this in (\ref{EC}),
one gets
\begin{equation}
E_C = (n+1)E - nTS.
\end{equation}
With (\ref{S}), we thus obtain
\begin{equation}
E_C = \frac{nV_n}{8\pi G}r_+^{n-1} \label{EC1}
\end{equation}
for the Schwarz\-schild black hole. Note that, due to $E_C = 2E$,
the Cardy-Verlinde formula (\ref{cardyverl}) would yield $S=0$,
which suggests that one has to use instead the generalization
(\ref{cardyverlmod}) of the usual Cardy formula (\ref{cardy}),
which is valid if the ground state has zero conformal weight.
As the only scale appearing in the metric (\ref{schwarz}) is
given by the Schwarz\-schild radius $r_+$, an obvious
generalization of (\ref{cardyverl}) would be
\begin{equation}
S = \frac{2\pi r_+}{n}\sqrt{E_C\cdot 2E}. \label{cardyverlflat}
\end{equation}
One easily verifies that the
Casimir energy (\ref{EC1}) and the thermodynamical quantities
(\ref{S}) satisfy this modified Cardy-Verlinde formula.\\

As a confirmation of (\ref{cardyverlflat}), we consider
the asymptotically flat Kerr black
holes \cite{Myers:1986un}\footnote{For simplicity we restrict
ourselves to the case of only one rotation parameter.}
\begin{eqnarray}
ds^2 &=& -\frac{\Delta_r}{\rho^2}[dt - a\sin^2\theta d\phi]^2
         + \frac{\rho^2}{\Delta_r}dr^2 + \rho^2 d\theta^2 \nonumber \\
     & & + \frac{\sin^2\theta}{\rho^2}[adt - (r^2+a^2)d\phi]^2
         + r^2\cos^2\theta d\Omega^2_{n-2}, \label{kerr}
\end{eqnarray}
where
\begin{eqnarray}
\Delta_r &=& (r^2 + a^2) - 2mr^{3-n}, \nonumber \\
\rho^2 &=& r^2 + a^2\cos^2\theta.
\end{eqnarray}
The inverse temperature, free energy, entropy, energy, angular momentum
and angular velocity of the horizon read \cite{Myers:1986un}
\begin{eqnarray}
\beta &=& \frac{4\pi(r_+^2+a^2)}{(n-1)r_+
          + \frac{(n-3)a^2}{r_+}}, \qquad
F = \frac{V_n}{16\pi G}r_+^{n-3}(r_+^2+a^2), \\
S &=& \frac{V_n}{4G}r_+^{n-2}(r_+^2+a^2), \qquad
E = \frac{nV_n}{16\pi G}r_+^{n-3}(r_+^2+a^2), \label{SEKerr} \\
J &=& \frac{aV_n}{8\pi G}r_+^{n-3}(r_+^2+a^2), \qquad
\Omega = \frac{a}{r_+^2+a^2},
\end{eqnarray}
where $r_+$ denotes the largest root of $\Delta_r=0$.
We can again define the Casimir energy as the violation of
the Euler identity, i.~e.
\begin{equation}
E_C = n(E - TS + pV - \Omega J).
\end{equation}
Using like above tracelessness of the stress tensor, $pV = E/n$,
we get
\begin{equation}
E_C = \frac{nV_n}{8\pi G}r_+^{n-3}(r_+^2+a^2) = 2E.
\end{equation}
This, together with the thermodynamical quantities (\ref{SEKerr}),
satisfies again the Cardy-Verlinde formula (\ref{cardyverlflat}).
If we write (\ref{cardyverlflat}) in the form (\ref{cardyverlmod})
with $L_0 = Er_+$ and $c/6 = 2E_Cr_+$, we see that the central
charge $c$ is equal to $6nS/\pi$. Monotonicity of this generalized
c-function with respect to temperature changes is thus equivalent
to positive specific heat, i.~e.~, to thermodynamic stability. Of
course for the Schwarz\-schild black hole (\ref{schwarz}) the
specific heat is always negative. For the Kerr solution
(\ref{kerr}) it is straightforward to show that the specific
heat $C_J = (\partial S/\partial T)_J$ is positive for $n=2$,
$1 < r_+^2/a^2 < 3+2\sqrt 3$, and for $n=3$, $0 < r_+^2/a^2 < 3$.
For $n \ge 4$ there are no regions of positive $C_J$.\\

As a final remark we would like to emphasize that in the
above considerations of asymptotically flat Schwarz\-schild and
Kerr black holes in $n+2$ dimensions, we explicitly assumed
tracelessness of the stress tensor of a hypothetical underlying field
theory in $n+1$ dimensions. We then found that a generalized Cardy-Verlinde
formula holds. This suggests that, like their cousins in AdS space,
also the black holes (\ref{schwarz}) and (\ref{kerr}) may
admit a dual description in terms of a
conformal field theory that lives in one dimension lower.\\

In this context it is interesting to note that the central charge
$c=6nS/\pi$ found above is proportional to $(r_+/l_P)^n$, where $l_P$
denotes the Planck length in $n+2$ dimensions. Due to the holographic
principle \cite{'tHooft:1993gx,Susskind:1995vu} there should be one
degree of freedom per
Planck volume, so $c$ represents the total number of degrees of freedom
on the event horizon, and thus makes indeed sense as a central charge.
Note that for the Schwarz\-schild black hole, one has
\begin{equation}
E = \frac c6 \pi^2 r_+ \left(\frac{2T}{n-1}\right)^2, \label{2dCFT}
\end{equation}
which is exactly the energy-temperature relation of a two-dimensional
CFT with characteristic length $r_+$ and effective temperature
$\tilde{T} = 2T/(n-1)$. Alternatively, one can write (\ref{2dCFT})
in the form
\begin{equation}
Er_+ = \frac{c}{24},
\end{equation}
which is the ground state energy of a CFT in two dimensions.

\section{Conclusions}

In the present paper we analyzed the thermodynamics of conformal
field theories with AdS duals, and showed that they share a
completely universal behavior, in that a generalized Cardy
formula is valid even when R-charges or bulk supergravity scalars
are turned on, or when the CFT resides on a rotating spacetime. 
We argued that the validity of the Cardy-Verlinde formula can
be traced back to the fact that these CFTs share many features in common
with conformal field theories in two dimensions. For instance,
they satisfy the general relation
\begin{equation}
2ER = \frac{n}{b\pi}S_C[1 + \Delta^{-2}],
\end{equation}
where the function $\Delta$ encodes the detailed properties
of the conformal field
theory, e.~g.~value of the coupling constant, presence of R-charges,
VEVs of certain operators or rotation.

Much of our intuition has been gained by the fact
that we were able to write the Casimir entropy $S_C$ like a
two-dimensional central charge. Then it was natural to think of $S_C$
as a generalized 
$C$-function. In this spirit we have studied the 
behavior of $S_C$ both under temperature changes and when non-trivial
bulk fields are turned on. The idea was to see if the presence of
non-trivial fields in the bulk theory were to induce a holographic
RG-flow in the dual field theory. The examples studied here support
such an 
interpretation. Indeed the Casimir entropy $S_C$, which in general is
a function of the temperature and of the renormalization scale of the
field theory,  
is a good candidate for a generalized $C$-function for field theories
that have a dual supergravity description, both at zero and at finite
temperature. We have found that $S_C$ is a monotonic function in each
variable, separately. 
Keeping the energy scale fixed we move in the space of conformally
invariant theories 
at various temperatures; if we fix the temperature and vary the energy
scale we move along the renormalization group trajectories. It becomes
extremely interesting to study the behavior of $S_C$ in complete
generality. 

For ${\cal N}=4$, $D=4$ SYM theory, one finds $b=2$ at strong coupling
and $b=4$ at weak coupling, suggesting that a generalized Cardy formula
may hold at all couplings, with $b$ a function of $g_{YM}^2N$,
interpolating smoothly (for temperatures above the Hawking-Page
phase transition) between the two regimes $g_{YM}^2N \to 0$ and
$g_{YM}^2N \to \infty$. It would also be of interest to test our
idea by explicitly calculating corrections to the Cardy-Verlinde
entropy formula both at weak and at strong coupling.

We did not succeed to show the validity of a generalized Cardy
formula for the $(0,2)$ free CFT in six dimensions or for the
${\cal N}=8$ supersingleton theory in three dimensions.
The fact that such
a formula holds for the weakly coupled $D=4$, ${\cal N}=4$ SYM
model, but not for the other CFTs with AdS duals (at weak coupling),
may be connected to the non-renormalization properties of the former
theory and deserves further study.

\section*{Acknowledgments}
\small

D.~K., A.~C.~P.~and D.~Z.~are partially supported by INFN, MURST and
by the European Commission RTN program
HPRN-CT-2000-00131, in which they are associated to the University of
Torino.
G.~S.~is supported by the US Department of Energy under grant
DE--FG05--91ER40627.

\end{document}